\def\bold#1{\setbox0=\hbox{$#1$}%
     \kern-.025em\copy0\kern-\wd0
     \kern.05em\copy0\kern-\wd0
     \kern-.025em\raise.0433em\box0 }
\def\slash#1{\setbox0=\hbox{$#1$}#1\hskip-\wd0\dimen0=5pt\advance
       \dimen0 by-\ht0\advance\dimen0 by\dp0\lower0.5\dimen0\hbox
         to\wd0{\hss\sl/\/\hss}}
\documentstyle[12pt]{article}

\newlength{\dinwidth}
\newlength{\dinmargin}
\newcommand{\resection}[1]{\setcounter{equation}{0}\section{#1}}

\begin{document}

\def\lq{\left [}
\def\rq{\right ]}
\def\LL{{\cal L}}
\def\VV{{\cal V}}
\def\AA{{\cal A}}

\newcommand{\be}{\begin{equation}}
\newcommand{\ee}{\end{equation}}
\newcommand{\bea}{\begin{eqnarray}}
\newcommand{\eea}{\end{eqnarray}}
\newcommand{\nn}{\nonumber}
\newcommand{\dd}{\displaystyle}

\thispagestyle{empty}
\vspace*{4cm}
\begin{center}
  \begin{Large}
  \begin{bf}
DISCONTINUITY THEOREM  FOR FIRST ORDER PHASE TRANSITIONS.
IMPLICATIONS FOR QCD$^*$\\
  \end{bf}
  \end{Large}
  \vspace{5mm}
  \begin{large}
A. Barducci, R. Casalbuoni and G. Pettini\\
  \end{large}
Dipartimento di Fisica, Univ. di Firenze\\
I.N.F.N., Sezione di Firenze\\
  \vspace{10mm}
  \begin{large}
R. Gatto\\
  \end{large}
D\'epartement de Physique Th\'eorique, Univ. de Gen\`eve\\
  \vspace{10mm}
\end{center}
  \vspace{1cm}
\begin{center}
UGVA-DPT 1992/12-798\\
   \vspace{3mm}
December 1992
\vspace{3mm}
hep-ph/9212276
\end{center}
\vspace{3cm}
\noindent
$^*$ Partially supported by the Swiss National Foundation

\newpage
\thispagestyle{empty}
\begin{quotation}
\vspace*{5cm}
\begin{center}
  \begin{Large}
  \begin{bf}
  ABSTRACT
  \end{bf}
  \end{Large}
\end{center}
  \vspace{5mm}
\noindent
A first order phase transition leading to deconfinement and chiral restoration
is a likely possibility for QCD, at least in some region of the
temperature-density plane. A signal for a unique
transition is that the order
parameters for such transitions (which can be understood in terms of symmetries
only in limiting situations of very massive or massless quarks) are both
discontinuous at the same critical temperature. We show that such a situation
can be understood on a precise thermodynamical basis because of a general
relation among discontinuities which holds for first order transitions. We
derive the result by a generalization of the Clausius-Clapeyron equation and
also through the effective action formalism. We illustrate the theorem in an
elementary example.
\end{quotation}
\newpage
\resection{Introduction and summary of conclusions}

There are two limiting situations in QCD where one expects the occurrence
of phase transitions characterized by corresponding symmetry breaking.
In the limit of infinite quark mass, the deconfinement transition can be
related
to the breaking of a
discrete group $Z(3)$, which is a symmetry of the QCD lagrangian
in that limit. The Polyakov loop acts as order
parameter for the transition. In the opposite limit, of massless quarks, the
chiral phase transition is related to restoration of chiral symmetry.
The fermionic condensate acts as the relevant order parameter.

In the realistic case of QCD with quarks of finite mass, neither the discrete
$Z(3)$ group, nor the global chiral symmetry are good symmetries of the
lagrangian. One may however still consider to look at what happens to the two
preceding order parameters, though they had been constructed for differing
limiting situations.

A computer simulation, with 4 light quark flavors with
dynamical mass $m/T\approx 0.1$, has shown an abrupt transition for the two
order parameters at practically the same temperature \cite{Karsch},
\cite{Satz}.

Another important issue is the order of the transition.  Recent results,
reviewed in ref. \cite {Peters}, point towards a first order phase
transition in the
case of two almost massless quarks and a massive one. It should also be
mentioned
that there appears to be a sort of critical value for the mass of the third
quark, above which the transition becomes  second order.

The problem of the transitions has also been studied in ref. \cite{Ellis}
assuming an effective lagrangian. The authors argue that the deconfined phase
should be characterized by the vanishing of the gluonic condensate,
$\langle 0|F_{\mu\nu}F^{\mu\nu}|0\rangle\beta(g)/(2g)$=0, and furthermore
that the deconfinement transition should be first order.
For this discussion it is
convenient to introduce separate critical temperatures $T_\chi$ and $T_d$
for the chiral
and the deconfinement transitions, as one
would expect when ignoring couplings between different degrees of freedom.
The authors of ref. \cite{Ellis} show that, if $T_\chi>T_d$, the
first order deconfinement transition drives a discontinuity of the
fermionic condensate. As a consequence the chiral transition appears
of first order, whichever was its nature when ignoring other degrees of
freedom,  and at the same temperature of the deconfinement one. We also mention
indications \cite{Gaust} that the topological susceptibility drives
the transition of other quantities, such
as the quark condensate or the Polyakov loop.

We shall prove here a general thermodynamical result. We shall show that
a generalized version of the Clausius-Clapeyron equation
relates discontinuities of the thermal averages of different observables
at the critical point of a first order phase transition. In a more
precise way, we consider a general hamiltonian, written in
the form $H=\sum_\alpha\lambda_\alpha\hat{\cal O}_\alpha$,
where $\lambda_\alpha$ are constants and $\hat{\cal O}_\alpha$
arbitrary operators.
The critical point of the first order phase transition will
depend, in general, on a subset of the parameters $\lambda_\alpha$,
to be called $\{\lambda_k\}$. Then
all the thermal averages of the
operators $\hat{\cal O}_k$,
related to the $\lambda_k$'s, are expected to become
discontinuous at the critical
point, both in their dependence from temperature and from density.

The consequence for QCD is obvious, assuming that there is a first order
transition and that the critical point depends on parameters, such as quark
masses,
the QCD coupling constant, etc. Then the quark condensate, the
gluon condensate, etc., are expected to exhibit simultaneous
discontinuities at such a critical point. Of course,
this is not a proof that such a scenario necessarily occurs in QCD, but it
clearly shows that such an apparently curious thing as a unique phase
transition, perceived through order parameters related to different, even
incompatible, limiting situations of the theory, has a simple and precise
thermodynamical basis.

For the derivation we first present a generalization of the Clausius-Clapeyron
equation (section 2). In section 3 we deal with the discontinuities. We give
a derivation based on the functional effective action formalism in section 4.
A most elementary example, developed in section 5, illustrates the simplicity
of the result.
\bigskip

\resection{Generalization of the Clausius-Clapeyron relation}

\def\SS{{\bf S}}
\def\BB{{\bf B}}
\def\CC{{\bf C}}
\def\AA{{\bf A}}
\def\ZZ{{\bf Z}}
\def\MM{{\bf M}}
\def\xx{{\bf x}}
\def\aa{{\bf a}}
\def\zz{{\bf z}}
\def\mm{{\bf m}}
\def\cc{{\bf c}}
\def\G{\Gamma_2}
\def \fiu {\varphi}
\def\GG{\Gamma}
\def\SG{{\bf \Sigma}}
\def\intpq{\int{{d^4p}\over{(2\pi)^4}}~{{d^4q}\over{(2\pi)^4}}}
\def\intq{\int{{d^4q}\over{(2\pi)^4}}}
\def\intp{\int{{d^4p}\over{(2\pi)^4}}}
\def\intl{\int^{\Lambda} {{d^4k}\over{(2\pi)^4}}}
\def\tr{{\rm tr}}
\def\Tr{{\rm Tr}}
\def\de{{\rm det}}
\def\Ps{\overline \Psi}
\def\OSS{\overline \SG}
\def\os{\overline \Sigma}
\def\trlog{{\tr}~ {\log}}
\def\Trlog{{\Tr}~ {\log}}
\def\trsp{{\rm tr}~({\bf s}^{2}+{\bf p}^{2})}
\def\trms{{\rm tr}~({\bf m}(\mu)\cdot {\bf s})}
\def \fp {f_{\pi}}
\def \mp {m_{\pi}}
\def \ss {\Sigma}
\def \psp {\langle \bar\psi \psi \rangle}
\def \b {\beta}
\def \sui { \sum_{n=-\infty}^{\infty}}
\def \su { \sum_{n=1}^{\infty}}
\def \suz {\sum_{i=1}^3}
\def \tc  {\left(1-{T\over T_c}\right)}
\def\prop{i \hat p - \sig (p^2)}
\def\fl{f_{\Lambda}}
\def \sp {\hskip3pt}
\def \dalamb {\sqcup \hskip-9pt \sqcap}
\def \O {{\cal O}}
\def \vi {\varphi}

To derive the relation we are interested in, let us
start by recalling  some thermodynamical relations.
Consider the partition function in the Grand
Canonical Ensemble
\be
Z={\rm Tr} ~ e^{-\beta ({\hat H}-\mu {\hat N})}
\ee
and a Hamiltonian of the form
\be
{\hat H}={\hat H}_{0}+\sum_{\alpha=1}^{N} ~ \lambda_{\alpha} ~ {\hat
\O}_{\alpha}
\ee
where $\lambda_{\alpha}$ are real parameters conjugate to the hermitean
operators
${\hat \O}_{\alpha}$.
The expectation value of the operator ${\hat \O}_{\alpha}$
is
\bea
\langle {\hat \O}_{\alpha}\rangle&=& - ~ {\partial ~ (T ~ {\rm ln} ~ Z)\over
\partial \lambda_{\alpha}}\nn\\
&=& {1\over Z} ~ {\rm Tr} ~ \Big[ {\hat \O}_{\alpha} ~
e^{-\beta({\hat H}-\mu {\hat N})}\Big]
\eea

In the same way, by varying with respect to the chemical
potential $\mu$, one has the mean particle number
(or it could be any conserved charge, and we are not necessarily
considering a fermionic theory)
\be
\langle {\hat N}\rangle= ~ {\partial ~ (T ~ {\rm ln} ~ Z)\over
\partial \mu}
\ee
We also  recall  that the entropy $S$ is
\be
S={\partial ~ (T ~ {\rm ln} ~ Z)\over \partial ~ T}
\ee
Now, let us assume that the theory describes a system which
can exist in two phases, each  described by an equation of state
$p_{i}=p_{i}(T,\mu) ~~ (i=1,2)$, where $p_{i}$ is the
pressure in the
particular phase; and that there exists a set of points
$(\mu_{c},T_{c})$ in the plane of chemical potential and
temperature, by crossing which the system
undergoes first order phase transitions.
In any of these points the equations of state of the two
phases coincide
\be
p_{1}(T_{c}(\{\lambda_{k}\}),\mu_{c} (\{\lambda_{k}\}),\{\lambda_{k}\})=
p_{2}(T_{c}(\{\lambda_{k}\}),\mu_{c} (\{\lambda_{k}\}),\{\lambda_{k}\})
\ee
($\{\lambda_{k}\}$ indicates the subset of parameters which affect
the critical behaviour).

\noindent
If the pressure depends smoothly on $\{\lambda_k\}$, we
can require the equality of the total derivatives of the two
members in eq. (2.6) with respect to $\lambda_{k}$
\be
{dp_{1}\over d \lambda_{k}}\Big|_{c}=
{dp_{2}\over d \lambda_{k}}\Big|_{c}
\ee
obtaining
\be
{\partial p_{1}\over\partial \lambda_{k}}\Big|_{c}
{}~ + ~ {\partial p_{1}\over\partial T}\Big|_{c}
{}~ {d T_{c}\over d  \lambda_{k}} ~ +
{}~ {\partial p_{1}\over\partial \mu}\Big|_{c}
{}~ {d \mu_{c}\over d \lambda_{k}}=
{\partial p_{2}\over\partial \lambda_{k}}\Big|_{c}
{}~ + ~ {\partial p_{2}\over\partial T}\Big|_{c}
{}~ {d T_{c}\over d \lambda_{k}} ~ +
{}~ {\partial p_{2}\over\partial \mu}\Big|_{c}
{}~ {d \mu_{c}\over d \lambda_{k}}
\ee
Here the subscript $\Big |_{c}$ means that the quantities
are evaluated at $(\mu_{c},T_{c})$.
We recall that according to the Ehrenfest classification,
in a first order phase transition the pressure
(or any another state function) is continuous
whereas its first (partial) derivatives are discontinuous.
Also, in the thermodynamic limit
\be
p ~ {\rm V}=T ~ {\rm ln} ~ Z
\ee
(here ${\rm V}$ is the three-dimensional volume and
the pressure has been taken vanishing at $T=\mu=0$)
Thus eq. (2.8) can be written in the form
\be
{\rm disc} ~ \langle {\hat o_{k}}\rangle=
 {d T_{c}\over d \lambda_{k}} ~ {\rm disc} ~ s
+ { d \mu_{c}\over d \lambda_{k}} ~ {\rm disc} ~ n
\ee
where
\be
\langle {\hat o_{k}}\rangle={\langle {\hat \O_{k}}\rangle\over
{\rm V}} ~ ~ ~ ~ ~ s={S\over {\rm V}} ~ ~ ~ ~ ~ n=
{\langle{\hat N}\rangle\over {\rm V}}
\ee
Let us now imagine two lines of first order critical points,
one at $\lambda_{k}$ and another at $\lambda_{k} + d\lambda_{k}$.
If we consider eq. (2.10) at $(\mu_{c},T_{c})$ on the line
at $\lambda_{k}$, we can unambiguously evaluate
the discontinuities of the densities
$\langle{\hat o}_{k}\rangle$, \sp $s$, \sp and \sp $n$,
whereas we have to specify on which point
$(\mu_{c}^{'},T_{c}^{'})$
of the line at $\lambda_{k} ~ + ~ d\lambda_{k}$
we end in order to calculate the two derivatives
$d T_{c}/ d \lambda_k$
and $d \mu_{c}/ d \lambda_{k}$.
In other words, although these derivatives are
bound to satisfy eq. (2.10), we have the freedom to
choose the direction along which to evaluate them.
In particular we can consider the variation
of $T_{c}$ (of $\mu_{c}$) with respect to $\lambda_{k}$
at fixed $\mu$  (at fixed $T$),
and in this way we can express
${\rm disc} ~ \langle {\hat o}_{k}\rangle$ only
in terms of of ${\rm disc} ~ s$
(of ${\rm disc} ~ n$), eliminating the other discontinuity.
Thus, under these conditions, we have the further relation
\be
{\partial T_{c}\over \partial \lambda_{k}}\Big|_{\mu}
{}~ {\rm disc} ~ s=
{\partial \mu_{c}\over \partial \lambda_{k}}\Big|_{T} ~ {\rm disc} ~ n
\ee
The previous result can be also expressed by the fact that,
in presence of a critical line, $T_{c}$ and $\mu_{c}$ are not
independent variables and that if we fix for instance
the chemical potential
$d T_{c}/d\lambda_{k}=\partial T_{c}/\partial \lambda_{k}$

Let us now consider the thermodynamic relation
\bea
\varepsilon &=& ~ - ~ p ~ + ~ T ~ {\partial p\over\partial T}
{}~ + ~ \mu ~ {\partial p\over\partial \mu}\nn\\
&=&- ~ p ~ + ~ T ~ s
{}~ + ~ \mu ~ n
\eea
which allows to determine the latent heat for a first
order phase transition
\be
{\rm disc} ~ \varepsilon ~ = ~ T_{c} ~ {\rm disc} ~ s ~ + ~ \mu_{c} ~
{\rm disc} ~ n
\ee
By using eqs. (2.10) and (2.12), ${\rm disc}~ \varepsilon$
can be entirely expressed
in terms of ${\rm disc} ~ \langle {\hat o}_{k}\rangle$
\be
{\rm disc} ~ \varepsilon=
\left( {1\over \partial ({\rm ln} T_{c})/ \partial \lambda_{k}
\Big|_{\mu}} ~ +
{}~ {1\over \partial ({\rm ln} \mu_{c})/ \partial \lambda_{k}
\Big|_{T}}\right)
{}~ {\rm disc} ~ \langle {\hat o}_{k}\rangle
\ee
The relation (2.15) is of the type of the Clausius-Clapeyron
equation for a liquid-vapor first order transition,
\be
{dT_{c}\over dp}={{\rm disc} ~ v\over {\rm disc} ~ s}
\ee
($v$ is the specific volume)
as shown in a recent paper by
H.Leutwyler \cite{Leutw}  for zero chemical potential.

However the relation (2.15) is quite more general; it
concerns any operator and it allows for a chemical potential.
We must remember that
to get (2.15) in this precise form, eq. (2.12) is necessary,
which requires the existence of lines of first order
critical points at varying parameters $\{\lambda\}$, rather
than an isolated critical point.
\bigskip
\resection{Relations among discontinuities}

Eqs. (2.10), (2.12) (and (2.15)) hold for the thermal
average  of any operator whose conjugate
parameter influences the critical point.
If ${\hat{\cal O}}_i$ and ${\hat{\cal O}}_j$ are two of such operators,
eliminating alternatively ${\rm disc} ~ s$
and ${\rm disc} ~ n$ from eqs. (2.10) and (2.12), written by
varying with respect to the parameters $\lambda_{i}$
and $\lambda_{j}$, we also get
\be
{\rm disc} ~ \langle {\hat o}_{i}\rangle=
{\partial T_{c}/\partial\lambda _{i}|_{\mu}\over
\partial T_{c}/\partial\lambda _{j}|_{\mu}} ~
{\rm disc} ~ \langle {\hat o}_{j}\rangle
\ee
\be
{\rm disc} ~ \langle {\hat o}_{i}\rangle=
{\partial \mu_{c}/\partial\lambda _{i}|_{T}\over
\partial \mu_{c}/\partial\lambda _{j}|_{T}} ~
{\rm disc} ~ \langle {\hat o}_{j}\rangle
\ee

Namely, all the operators conjugate to parameters which
affect the critical point are expected to be discontinuous at the
phase transition point. The ratio of their discontinuities is
simply the ratio of the variation of the critical temperature
or chemical potential with respect to the correspondent
conjugate parameters.
{}From the two previous equations one also gets
\be
{\partial T_{c}/\partial\lambda _{i}|_{\mu}\over
\partial T_{c}/\partial\lambda _{j}|_{\mu}} ~ =
{\partial \mu_{c}/\partial\lambda _{i}|_{T}\over
\partial \mu_{c}/\partial\lambda _{j}|_{T}} ~
\ee

Of course we could have employed other
variables instead of the pressure
and followed the procedure starting from eq. (2.6)
obtaining relations similar to eq. (2.15).
For instance, by equating the chemical potentials $\mu_{i} (p,T)$
in the two phases for the liquid-vapor transition, varying
with respect to $p$ ($T_{c}$ is a function of $p$),
one obtains eq. (2.16).
The reason why we have chosen the pressure will be clearer
in the following.

In this context our aim is to stress the essential ingredients.
If in some system a first order phase transition
occurs, using the fact that the state functions
are continuous, but not their
partial derivatives, one can easily relate the various discontinuities
appearing in the description, obtaining alternative expressions for
quantities such as the latent heat.
In particular these relations can be useful,
as already remarked by Leutwyler \cite{Leutw}, if one can easier compute the
discontinuity of some operator rather than the discontinuities
of the derivatives of the pressure.

The previous analysis can readily be extended to cases where there is a set
of chemical potentials $\mu_{i}$ coupled to charge operators
${\hat N}_{i}$.
\bigskip

\resection{The effective action derivation}

Let us consider a field theory at finite temperature $T$ and
chemical potential $\mu$, described by
an euclidean effective action $\GG$. The effective action is a functional of
a set of operators $\{ {\hat A}\}$,
the dynamical variables of the theory,
\be
\GG=\GG \Big[\{{\rm \hat A}\};\{\alpha \};T,\mu\Big]
\ee
In the previous equation we have also explicitly written the dependence
of $\GG$ on a set of parameters $\{\alpha\}$, which are
external sources conjugate to the operators $\{ {\hat A}\}$.
Suppose that by varying the temperature $T$ and/or the
chemical potential $\mu$, the effective action can describe
different phases of the theory, and that there exist lines of
first order critical points $(T_{c},\mu_{c})$ in the $(T,\mu)$ plane.
Then in any of these points $(T_{c},\mu_{c})$,
$\GG$ has two degenerate absolute minima $\{{\hat A}\}_{1}$,
$\{{\hat A}\}_{2}$, where we can write
the condition
\be
\GG\Big[\{{\hat A}\}_{1};\{\alpha\};T_{c}(\{\alpha\}),
\mu_{c}(\{\alpha\})\Big]=
\GG\Big[\{{\hat A}\}_{2};\{\alpha\};T_{c}(\{\alpha\}),
\mu_{c}(\{\alpha\})\Big]
\ee
Then we equate the total derivatives with respect to any
parameter $\alpha_{k}$ which affects the critical behaviour,
getting, apart from terms continuous at the transition,
\be
\Big[{\partial \GG\over \partial \alpha_{k}} ~ + ~
{\partial\GG\over \partial T}{d T_{c}\over d\alpha_{k}} ~
+ ~ {\partial\GG\over \partial \mu}{d \mu_{c}\over d\alpha_{k}} ~
+ ~ {\partial\GG\over\partial\{\hat A\}}{d\{\hat A\}\over d\alpha_{k}}\Big]
\Big|_{c}^{\{A\}_{1}}= ({\rm same})~\Big|_{c}^{\{A\}_{2}}
\ee
The derivatives of $\GG$ with respect to
$\{\hat A\}$ vanish on  both sides of the previous
expression, since they are evaluated at
the two minima of the effective action.
Actually the effective action evaluated at these points
is nothing but $-{\rm ln}Z$ and therefore
$~-p~V/T$ (since it is defined for infinite volume)
for the two coexisting phases. Thus at the
physical point it looses its dependence on the dynamical
variables, whereas it remains a function of the sources
$\{\alpha\}$. Furthermore, in the case of constant fields,
we can employ the euclidean effective potential
\be
{\cal V}={\Gamma\over \beta V}=-p
\ee
which directly gives the pressure.

It follows that eq. (4.3), written for the effective
potential, is equivalent to eq. (2.8).

\bigskip

\resection{A simple example}

Let us now consider a simple example.
It is almost trivial, as it involves a single field. However it
 is useful to understand how the theorem works.

Consider an effective potential containing powers up to the sixth order of
a scalar field $\phi$, which plays the role of an order parameter \cite{Huang}:
\be
{\cal V}= a\phi^2-b\phi^4+c\phi^6 +h\phi
\ee
where we assume $c>0$ and $h$ small. We will show how the equations
(2.10), (2.15) are satisfied in this particular case, by
taking $h$ as the parameter $\lambda$ to be varied. Of course we could
use any of the other parameters appearing in eq. (5.1) but we will
limit ourselves to this case. At the end of the section we will discuss also
eq. (3.1) and then we will also use the other parameters appearing in
eq. (5.1) as variable parameters $\lambda_i$.
We start by discussing the model at $h=0$.
The potential has a critical line of second-order transitions for $b<0$ and
$a=0$. At $a=b=0$ the model has a tricritical point \cite{Griff}, whereas
for $b>0$ there is a line of first-order transitions given by the equation
\be
a=\frac{b^2}{4c}
\ee
The tricritical point is where the two lines of first- and second-order
transitions merge together. At any of the first-order transition
points one has two phases characterized by the following values of $\phi$
(interpreted as the vacuum expectation values, v.e.v.)
\be
\langle\phi\rangle_1=0,~~~~\langle\phi\rangle_2=
\sqrt{\frac{b}{2c}}
\ee
We also define the discontinuity as
\be
{\rm disc}~\langle\phi\rangle=\langle\phi\rangle_2-\langle\phi\rangle_1=
\sqrt{\frac{b}{2c}}
\ee
For the following discussion we will assume also a linear dependence of the
parameter $a$ on the temperature and on the chemical potential around the
first-order critical line. We could assume an analogous dependence for any
of the parameters, but to keep the matter simple we will stay with $a$.
We put
\be
a=\alpha T+\gamma \mu-a_0
\ee
obtaining for the critical line
\be
\alpha T_c+\gamma\mu_c=a_0+\frac{b^2}{4c}
\ee
When we insert the small perturbation $h\phi$ we need (at the lowest order in
$h$) only to know how the critical quantities are changed
\be
\alpha T_c+\gamma\mu_c=a_0+\frac{b^2}{4c}-\sqrt{\frac{2c}{b}}h
\ee
By noticing that in this case the pressure is given by $p=-{\cal V}$, we get
for the specific entropy and for the density
\be
s=-\alpha\langle\phi^2\rangle,~~~~n=-\gamma\langle\phi^2\rangle
\ee
By taking $\lambda=h$ we can verify eq. (2.10)
\bea
&&\frac{dT_c}{dh}~{\rm disc}~s+\frac{
d\mu_c}{dh}~{\rm disc}~n=\nn\\
&=&
\left(-\frac{1}{\alpha}\sqrt{\frac{2c}{b}}-\frac{\gamma}{\alpha}\frac{
d\mu_c}{dh}
\right)\left(-\alpha~{\rm disc}~\langle\phi^2\rangle\right)+
\frac{d\mu_c}{dh}
\left(-\gamma~{\rm disc}~\langle\phi^2\rangle\right)=\nn\\
&=&\sqrt{\frac{2c}{b}}\frac{b}{2c}=
{}~{\rm disc}~\langle\phi\rangle
\eea
{}From eq. (5.8) and (2.14) we can derive the discontinuity for the energy
density
\be
{\rm disc}~\epsilon=(-\alpha T_c-\gamma\mu_c)~{\rm disc}~\langle\phi^2\rangle
\ee
Then, by differentiating (5.7)
\be
\frac{\partial T_c}{\partial h}\Big |_\mu=
-\frac{1}{\alpha~{\rm disc}~\langle\phi\rangle}~,~~~~~~~~~~~~~~~~
\frac{\partial\mu_c}{\partial h}\Big |_T=
-\frac{1}{\gamma~{\rm disc}~\langle\phi\rangle}
\ee
it is simple matter to verify eq. (2.15). Finally we verify eq. (3.1) by
taking $h$ and $c$ as $\lambda$ parameters. Using
\be
{\rm disc}~\langle\phi^6\rangle=\left(\frac{b}{2c}\right)^3
\ee
we get (in the limit $h\to 0$)
\bea
{\rm disc}~\langle\phi^6\rangle&=&\frac{\partial T_c/\partial c\Big |_\mu}
{\partial T_c/\partial h\Big |_\mu}~{\rm disc}~\langle\phi\rangle=\nn\\
&=&
\left(-\frac{1}{\alpha}\frac{b^2}{4c^2}\right)\left(-\frac{1}{\alpha}
\sqrt{\frac{2c}{b}}\right)^{-1}\sqrt{\frac{b}{2c}}=\left(\frac{b}{2c}\right)^3
\eea
All the other relations can be proved in analogous way.


\newpage

\begin{thebibliography}{99}

\bibitem{Karsch}
F.Karsch, J.Kogut, D.K.Sinclair and H.W.Wyld, Phys. Letters {\bf 188B}
(1987) 353

\bibitem{Satz}
For a review see:
H.Satz, Proceedings of the Aachen Workshop on LHC, Aachen, 4-9 October 1990,
CERN 90-10, ECFA 90-133, Eds. G.Jarlskog and D.Rein, Vol. 1, p. 188

\bibitem{Peters}
B.Petersson, Proceedings of the Eight International Conference on
Ultrarelativistic Nucleus-Nucleus Collisions, Quark Matter 1990,
Menton, France, May 7-11, 1990, Eds. J.P.Blaizot, C.Garschel, B.Pire and
A.Romana, Nucl. Physics {\bf A525} (1991) 237c

\bibitem{Ellis}

B.Campbell, J.Ellis and K.A.Olive, Nucl. Physics {\bf B345} (1990) 57

\bibitem{Gaust}

H.Gausterer, J.Potvin and S.Salenievici, Phys. Review {\bf D41} (1990) 3829

\bibitem{Leutw}

H.Leutwyler, Phys. Letters {\bf 284B} (1992) 106

\bibitem{Huang}

For a discussion of this model see: K.Huang, "Statistical Mechanics", 2$^{nd}$
edition, edited by J.Wiley and Sons, p. 428 (1987)

\bibitem{Griff}

We follow the terminology suggested by R.B.Griffiths, Phys. Rev. Letters
{\bf 24} (1970) 715

\end{thebibliography}
\end{document}